\documentclass[twocolumn, tighten, times]{aastex63}
\usepackage[utf8]{inputenc}
\usepackage{bm}
\usepackage{tabularx}
\usepackage{braket}
\usepackage{multirow}
\usepackage{xcolor}
\usepackage{pifont}
\usepackage{enumitem}
\usepackage{soul}
\usepackage{xcolor}

\shortauthors{Lin et al.}

\graphicspath{{./}{figures/}}

\begin{document}

\title{Investigating the Hubble Constant Tension - Two Numbers in the Standard Cosmological Model}

\author[0000-0003-2240-7031]{Weikang Lin}
\email{wlin23@ncsu.edu}
\affiliation{Physics Department, North Carolina State University, Raleigh, NC 27695, USA}
\author[0000-0001-8927-1795]{Katherine J. Mack}
\affiliation{Physics Department, North Carolina State University, Raleigh, NC 27695, USA}
\affiliation{Perimeter Institute for Theoretical Physics, Waterloo ON N2L 2Y5, Canada}
\author{Liqiang Hou}
\affiliation{Physics Department, North Carolina State University, Raleigh, NC 27695, USA}

\begin{abstract}
The current Hubble constant tension is usually presented by comparing constraints on $H_0$ only. However, the post-recombination background cosmic evolution is determined by two parameters in the standard $\Lambda$CDM model, the Hubble constant ($H_0$) and today's matter energy fraction ($\Omega_{\rm{m}}$). If we therefore compare all constraints individually in the $H_0$-$\Omega_{\rm{m}}$ plane, (1) various constraints can be treated as independently as possible, (2) single-sided constraints are easier to consider, (3) compatibility among different constraints can be viewed in a more robust way, (4) the model dependence of each constraint is clear, and (5) whether or not a nonstandard model is able to reconcile all constraints in tension can be seen more effectively. We perform a systematic comparison of different constraints in the $H_0$-$\Omega_{\rm{m}}$ space based on a flat $\Lambda$CDM model, treating them as separately as possible. Constraints along different degeneracy directions consistently overlap in one region of the space, with the local measurement from Cepheid variable-calibrated supernovae being the most outlying, followed by the time-delay strong-lensing result. Considering the possibility that some nonstandard physics may reconcile the constraints, we provide a general discussion on nonstandard models with modifications at high, mid, or low redshifts, and the effect of local environmental factors. Due to the different responses of individual constraints to a modified model, it is not easy for nonstandard models to reconcile all constraints if none of them have unaccounted-for systematic effects.
\end{abstract}


\section{Introduction} \label{sec:intro}
Cosmology used to be called ``a search for two numbers,'' referring to the Hubble constant and the deceleration parameter \citep{Sandage-1970-twonumbers}. While the former describes today's cosmic expansion rate and, when first discovered, caused Einstein to abandon his idea of the cosmological constant $\Lambda$, the latter turned out to be negative and prompted physicists to bring $\Lambda$ back. Dubbed (flat) ``$\Lambda$CDM,'' the simplest cosmological model fully determines the dynamics of the homogeneous universe with another combination of two numbers, this time pairing the Hubble constant $H_0$ with today's matter energy fraction $\Omega_{\rm{m}}$. Together with its description of large-scale inhomogeneities, this model has successfully explained various cosmological and astronomical observations. Its simplicity and (at least overall) concordance has made it the standard cosmological model, even while named after its two most mysterious aspects. The two numbers, $H_0$ and $\Omega_{\rm{m}}$, are the focus of this work.

Despite its successes, some tensions have recently been reported between observations based on the standard cosmological model. Among them, the Hubble constant tension is one of the most hotly debated: the local determination of $H_0$ based on Cepheid variable-calibrated Type Ia supernovae (SNe) \citep{Riess-etal-2019} is higher than the one inferred from cosmic-microwave-background (CMB) observations \citep{2018-Planck-cosmo-params} at a $4.4$-$\sigma$ confidence level, a discrepancy that has kept increasing with more precise data from both sides in the past decade. For the time being, it is unclear whether this tension is caused by some new physics beyond the standard cosmological model, or some systematic effects in either or both of the measurements. 

Looking for other independent observations is important and pressing, as this can help us to draw a more robust conclusion on the cause of the $H_0$ tension. Unfortunately, most other constraints on $H_0$ are not currently precise enough to settle the question, and their model dependences make the comparison more difficult to interpret. Nonetheless, $\Lambda$CDM can be taken as the default model to which all others can be compared. And because there are two parameters ($H_0$ and $\Omega_{\rm{m}}$) that specify the background evolution in $\Lambda$CDM, we should not compare constraints only on $H_0$. It is more instructive to perform a comparison in the $H_0$-$\Omega_{\rm{m}}$ space. While other authors sometimes perform this comparison, the benefits of doing so have rarely been discussed in the literature.

Usually, in order to obtain stronger constraints on $H_0$, different observations are combined to break degeneracies. Doing this not only reduces the number of constraints to compare, but also causes the joint results to correlate with each other as certain constraints are frequently used to break degeneracies (e.g., SNe). The model dependence of a joint result gets more complicated as well, because a new model may change only some constraints in a joint study but not the others. We will take a different approach and treat constraints as separately as possible - each constraint is obtained using a minimal number of observations (see the Appendix \ref{appendix:remarks-BAO-BBN} for a discussion about baryon acoustic oscillations) and we them individually in the $H_0$-$\Omega_{\rm{m}}$ plane. In the $H_0$-$\Omega_{\rm{m}}$ plane, several such constraints are actually not weak compared to the local determination, because their favored parameter spaces are relatively small. Therefore, it is not necessary for constraints to be strong on $H_0$ alone in order to be included in the comparison. Comparing constraints in such a way makes it clear which observations, when combined, can break degeneracies to give stronger constraints on $H_0$, and to see if those combinations would push us to unacceptable $(H_0, \Omega_{\rm{m}})$ regions. Considering observations individually also allows us more clearly see which constraint in the $H_0$-$\Omega_{\rm{m}}$ plane would be altered in a nonstandard model, providing information on whether a proposed model can reconcile all constraints and better clarifying the model dependence of each constraint, as advocated by \citet{Verde-etal-2019}. In addition, general single-sided constraints are easier to include in the $H_0$-$\Omega_{\rm{m}}$ plane. We will demonstrate the advantages of this approach here and investigate the current $H_0$ tension with a thorough comparison of different constraints. We then discuss whether any nonstandard models are capable of reconciling all constraints, assuming systematic effects are not biasing any results.
\begin{figure*}[tpb]
    \centering
    \includegraphics[width=\textwidth]{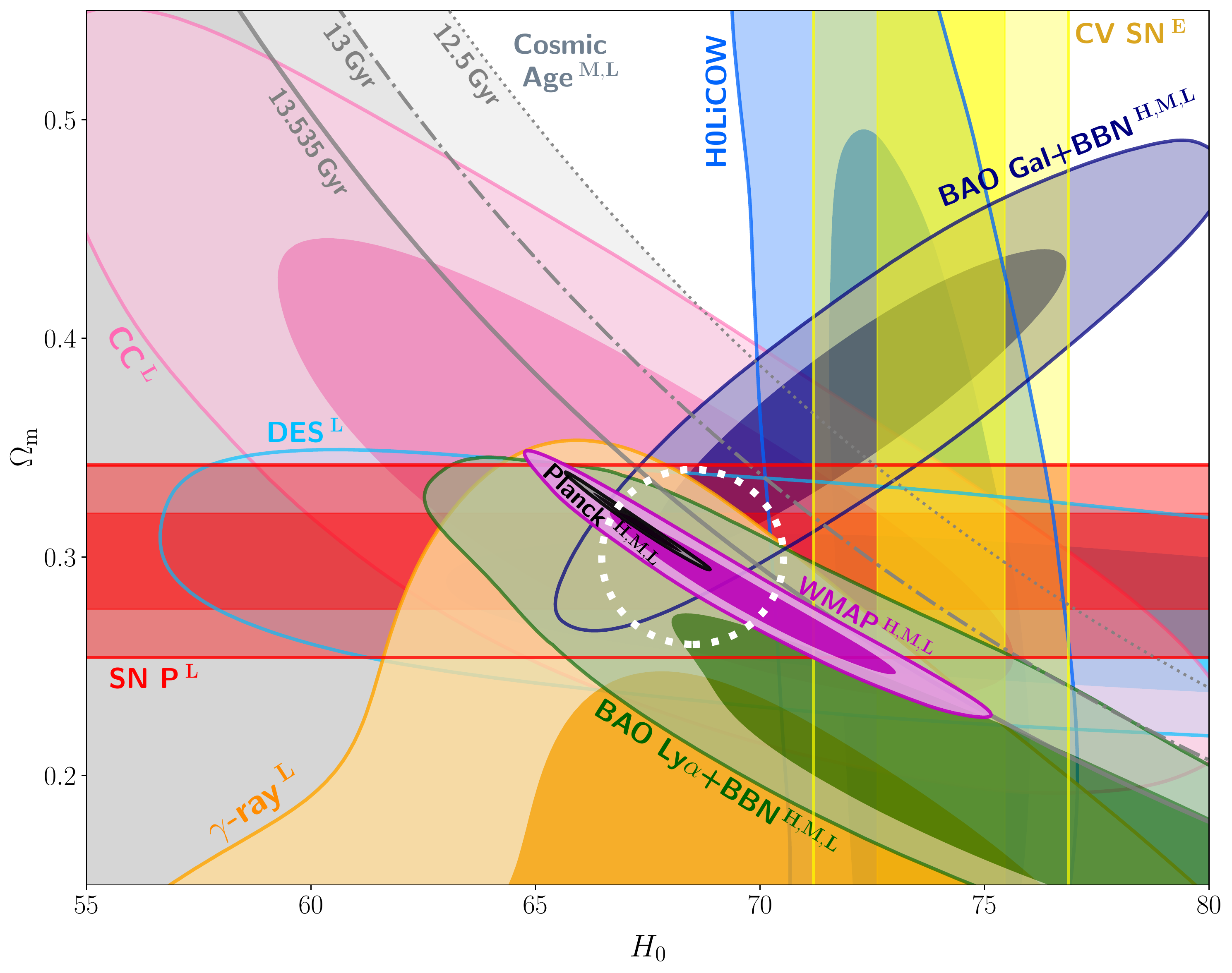}\vspace{-7pt}
    \caption{Different constraints in the $H_0$ and $\Omega_{\rm{m}}$ space based on a flat $\Lambda$CDM model. Dark and light contours show the $68\%$ and $95\%$ confidence regions of each posterior, with the exception of the cosmic-age bounds. (For each of those, parameter space outside the orange regions is excluded if the Universe is at least the age given in the label.) Most constraints with different degeneracy directions consistently overlap on the region indicated by the guiding white and dashed circle. Note that the circle \emph{does not} represent a joint constraint. Such a common region is however not overlapped by the Cepheid-based local determination of $H_0$ (CV SN) and is only marginally overlapped by the H0LiCOW constraint.  \\
    Contours correspond to SN P (red), DES (light blue), CC (pink), H0LiCOW (blue), BAO Gal (navy), BAO Ly$\alpha$ (green), $\gamma$-ray (orange), WMAP (magenta), Planck (black), CV SN (yellow) and some guiding cosmic-age constraints ($t_*=13.535$, $13$ and $12.5$\,Gyr; orange). See the text for descriptions and sources of those constraints. Each constraint in the figure is labeled according to whether it can be changed by nonstandard high-$z$ models (H), mid-$z$ models (M), low-$z$ models (L), or local environmental factors (E). See the text for the definition of those model categories. We leave the H0LiCOW technique without a label because it is relatively insensitive to the underlying cosmological model.}
    \label{fig:all-H0-Om-constraints}
\end{figure*}

\section{Methods and results}\label{sec:methods-results}
In order to perform a throughout multi-constraint comparison, we have collected a number of independent constraints in the $H_0$-$\Omega_{\rm{m}}$ space obtained from different observations, assuming the standard flat $\Lambda$CDM model. We list them below.
\begin{description}[align=left,leftmargin=8pt]
\setlength{\itemsep}{0pt}%
\setlength{\parskip}{0pt}%
\item[Time-delay strong lensing (H0LiCOW)] the time-delay distance for strong lensing systems measures $H_0$ in a way that is insensitive to the underlying cosmological model. The H0LiCOW constraint is obtained by running the jupyter notebook provided in \href{www.h0licow.org}{www.h0licow.org} \citep{Wong-etal-2019-joint-stronglensing} which includes six lens systems. We also added to the notebook a seventh lens system [included in \citep{Shajib-etal-2020} not publicly released] according to the description in Appendix \ref{appendix:lens7-likelihood}. In our numerical analysis, we also consider the recent result from \citep{Birrer-2020} as an alternative constraint (TDCOSMO), which relaxed the strong assumption on the lens model as used in \citet{Wong-etal-2019-joint-stronglensing}.

\item[Type-Ia supernovae Pantheon (SN P)] the relative change of the apparent magnitude of these standard candles as a function of redshift tightly constrains $\Omega_{\rm{m}}$ \citep{Scolnic-etal-2018};
\item[$\gamma$-ray attenuation ($\gamma$-ray)] the optical depth along the light of sight depends on the cosmic evolution \citep[chains obtained from private conversation]{Dominguez-etal-2019};

\item[Dark Energy Survey Year 1 (DES)] correlations among galaxies and cosmic shear (3$\times$2 correlation functions) \citep[chains obtained from \href{https://des.ncsa.illinois.edu}{https://des.ncsa.illinois.edu}]{DES-year1-2018-3x2}  (as most other large-scale structure (LSS) results are fairly uninformative in the $H_0$-$\Omega_{\rm{m}}$ plane, we only include the DES constraint here);

\item[Other large-scale-structure (LSS) constraints] Most other LSS constraints on the $H_0$-$\Omega_{\rm{m}}$ plane are week and prior limited. Here we consider two more constraints on this plane: galaxy clusters from the SPT-SZ survey (SPT SZ) \citep[chains obtained by using the package from \href{https://pole.uchicago.edu/public/data/sptsz-clusters/}{https://pole.uchicago.edu}]{Bocquet-etal-2019} and relaxed galaxy clusters \citep[a constraint on $\Omega_{\rm{m}}h^{1/2}$ in their table 5]{Mantz-etal-2014}. We include both constraints in our numerical analysis but only plot SPT SZ in Figure \ref{fig:other-localH0} in Appendix \ref{appendix:otherlocal} for clarity.

\item[Cosmic chronometers (CC)] the differential ages of passively evolving galaxies at two nearby redshifts directly measure the cosmic expansion rate at that redshift. The constraint here is obtained by performing a likelihood analysis of the data compilation discussed in \citet{2017-Moresco-Marulli-CC}.  We have added estimated systematic errors considered in \citet{2020-Moresco-Raul-Verde}; more details in Appendix \ref{appendix:additional-sys-cc};

\item[Galaxy BAO + BBN (BAO Gal)] galaxy baryon acoustic oscillations at $z_{\rm{eff}}=0.106$ \citep{Beutler-etal-2011}, $0.15$ \citep{Ross-etal-2015}, $0.38$, $0.51$, $0.61$ \citep{Alam-etal-2017} and $1.52$ \citep{Ata-etal-2018};

\item[Lyman-$\alpha$ BAO + BBN (BAO Ly$\alpha$)] baryon acoustic oscillations from Ly$\alpha$ auto correlation at $z=2.34$ and Ly$\alpha$-quasar cross correlation at $z=2.35$ \citep{Blomqvist-etal-2019}. The two BAO constraints are obtained by running the corresponding modules in \textsc{CosmoMC} \citep{Cosmomc-Lewis-Bridle-2002} with the big bang nucleosynthesis (BBN) constraint on $\Omega_{b}h^2=0.0222\pm0.0005$ \citep{Cooke-etal-2018};

\item[WMAP 2013 (WMAP)]  9-year Wilkinson Microwave Anisotropy Probe CMB data \citep{WMAP-year9-cosmological-param};

\item[Planck 2018 (Planck)]  full-mission Planck baseline CMB temperature and polarization data \citep{2018-Planck-cosmo-params}. Chains of both WMAP and Planck are obtained from \href{https://pla.esac.esa.int}{https://pla.esac.esa.int});

\item[Cepheid-calibrated SNIa (CV SN)] local determination of $H_0$ using Type Ia SNe calibrated by Cepheid variables \citep{Riess-etal-2019}. There is a small correlation between the SN P constraint on $\Omega_{\rm{m}}$ and the local measurement on $H_0$, which we ignore; also see \cite{2020-Dhawan-etal} for discussions.

\item[Alternative Local Measurements] For the purpose of our numerical analysis, We also consider the result from the megamaser cosmology project (MCP) \citep{Pesce-etal-2020} as an independent local measurement, as well as the Tip-of-the-Red-Giant-Branch (TRGB) result from \citet{Freedman-etal-2019} as an alternative distance ladder measurement, but for clarity we do not include these in Figure~\ref{fig:all-H0-Om-constraints} but show them individually in Figure~\ref{fig:other-localH0} in Appendix \ref{appendix:otherlocal}.

\item[Cosmic age] the cosmic time since $z=100$ ($t_{\rm{age}}^{100}$) should be larger than some estimated stellar ages ($t_*$)\footnote{We conservatively assume no stars could have formed before $z=100$. A larger limiting redshift does not change the bounds significantly. But a larger confirmed stellar age or a later stellar formation time will tighten the bounds towards a smaller-$H_0$ or smaller-$\Omega_{\rm{m}}$ direction.}. For example, based on a HST FGS parallax, the age of HD 140283 is estimated to be $t_*=14.27\pm0.8$\,Gyr \citep{2014_VandenBerg_etal_HD_140283}, though this stellar age was recently re-estimated to be $13.5\pm0.7$\,Gyr using the Gaia DR2 parallax \citep{2019-Jimenez-etal}. In addition, \citet{Schlaufman-etal-2018} estimated J18082002-5104378 A to have an age of $13.535\pm0.002$\,Gyr based on the Dartmouth isochrone library, though this only includes a statistical error. To investigate possible systematic effects, \citet{Schlaufman-etal-2018} considered two additional isochrone libraries, which gave lower estimates. However, they still prefer the higher estimate, since the Dartmouth library accounts for the $\alpha$-enhanced composition and has a better fit to the data \citep{Schlaufman-etal-2018}. More work is needed to arrive at a robust error estimate. Some Galactic globular clusters also have high estimated ages, e.g., $13.4\pm1.3$\,Gyr for NGC 5466 and $13.4\pm1.5$\,Gyr for NGC 2298, NGC 6101 and NGC 6341 \citep[Table 6]{O_Malley-etal-2017}. To show how stellar ages put constraints on the $H_0$-$\Omega_{\rm{m}}$ plane, we plot three guiding allowed regions (orange) in Figure~\ref{fig:all-H0-Om-constraints} using the estimated age of J18082002-5104378 A (if confirmed) as well as $t_{\rm{age}}^{100}>13$\,Gyr and $12.5$\,Gyr. Although the uncertainties are large for most of stellar ages quoted here, their mean values are consistently higher than the $t_{\rm{age}}^{100}\simeq12.7$\,Gyr suggested by a $\Lambda$CDM universe with $H_0\simeq74$\,km/s/Mpc and $\Omega_{\rm{m}}\simeq0.3$.  
\end{description}

We present all the above constraints in Figure~\ref{fig:all-H0-Om-constraints}, except the alternative local measurements. We omit these because the uncertainty of the MCP result is relatively large compared to CV SN and we consider the TRGB-based result (which is consistent with most other constraints) as an alternative to CV SN; we do, however, consider them in our numerical analysis and plot them individually in Figure~\ref{fig:other-localH0} in Appendix \ref{appendix:otherlocal}.

To check the consistency of a number of constraints, it is instructive to consider whether there is some common parameter region simultaneously overlapped by them. Note that observations only need to give some favored region in the $H_0$-$\Omega_{\rm{m}}$ space; they may not necessarily give strong constraints on $H_0$. For example, SN P is uninformative on $H_0$, and the stellar age constraint is only single-sided. Among those constraints, while both the local measurement and the H0LiCOW technique are insensitive to a cosmological model, other constraints are more model dependent. However, being model insensitive does not mean being free from systematic effects, which may affect any result. We therefore first analyze such a multi-constraint comparison based on the standard cosmology taking the face values of those constraints reported in the literature. Possible unaccounted-for errors for particular observations and nonstandard cosmological models will be discussed later. 

We can see in Figure~\ref{fig:all-H0-Om-constraints} that most constraints overlap at a common parameter space indicated by the guiding white and dashed circle. We stress that this circle is a guide, \emph{not} a joint constraint. Importantly, the overlapping constraints are along different degeneracy directions, which more robustly shows that those constraints are consistent with each other. This point cannot be seen from a comparison of constraints on $H_0$ alone. In the standard $\Lambda$CDM model, two constraints, CV SN and H0LiCOW, are noticeably incompatible with the other constraints. Their overlap with other constraints occur at different parameter regions, depending on which constraints we wish to reconcile. Roughly speaking, they overlap best with BAO Gal at $\Omega_{\rm{m}}\sim0.4$; with SN P and DES at $\Omega_{\rm{m}}\sim0.3$; with BAO Ly$\alpha$, $\gamma$-ray attenuation and WMAP at $\Omega_{\rm{m}}\sim0.23$.  If a star is confirmed to be older than $13$ Gyr, CV SN and H0LiCOW would overlap with the cosmic age constraint at $\Omega_{\rm{m}}\lesssim0.27$, and at a even lower range with a higher stellar age.

\begin{table}[tpb]
    \scriptsize
    \centering
    \caption{Multi-dataset IOIs and $\mathcal{O}_j$'s for our considered constraints (not including the cosmic-age bounds). If one constraint has a significantly higher $\mathcal{O}_j$ than the others, it is considered outlying. The first row is the analysis of all constraints; the MCP result is included in the analysis. The second to the fifth rows are those after removing CV SN, CV SN and H0LiCOW, WMAP, Planck, respectively. We replace CV SN with TRGB in the sixth row, and replace H0LiCOW with TDCOSMO in the seventh. In the last row, we combine the two BAO results. }
    \label{tab:IOI-and-outlier-index}
    \begin{tabular}{r|rrrrr}
    \hline\hline
     (IOI) \& $\mathcal{O}_j$  & CV SN & Planck & H0LiCOW & BAO Gal  & Others       \\
    \hline
    All (3.71)&  6.74 & 4.22  & 4.22 & 2.41 & $<$1.8\\
    \st{CV SN} (2.81) & N/A & 1.97 & 5.12 & 1.43 & $<$2.0 \\
    \st{CV SN, H0LiCOW} (2.08) & N/A & 0.80 & N/A & 1.62 & $<$2.1\\
    \st{WMAP} (3.71) & 6.75 & 5.28 & 4.22 & 2.41 & $<$1.8 \\
    \st{Planck} (3.07) & 3.96 & N/A & 2.19 & 4.31 & $<$1.7 \\
    TRGB \st{CV SN} (2.62) & N/A & 2.13 & 4.99 & 1.42 & $<$1.9\\
    TDCOSMO \st{H0LiCOW} (2.77) & 7.66 & 2.41 & N/A & 1.55 & $<$1.9 \\ 
    Combined BAO (3.24) & 6.87 & 3.87 & 2.94 & N/A & $<$1.8\\
    \hline
    \end{tabular}
\end{table}

It is useful to find a numerical way to quantify and generalize the above multi-constraint comparison. To do that we use the momentum-based multi-dataset Index of Inconsistency (IOI) \citep{WL2017a} and the associated ``outlier index'' ($\mathcal{O}_j$) \citep{Lin-Ishak-2019},
\begin{equation}\label{eq:outlierness-definition}
    \mathcal{O}_j\equiv \frac{1}{2}\big[N_{\rm{d}}{\rm{IOI}}-(N_{\rm{d}}-1){\rm{IOI}}^{(j)}\big]-\frac{N_{\rm{p}}-1}{2}\,,
\end{equation}
where $N_{\rm{d}}$ is the number of constraints, $N_{\rm{p}}$ is the number of parameters and IOI$^{(j)}$ is the multi-dataset IOI for the ($N_{\rm{d}}$-1) constraints excluding the $j$th one. Given a set of constraints, IOI quantifies their overall inconsistency and $\mathcal{O}_j$ tells us how incompatible each constraint is with the others. We show the results in Table \ref{tab:IOI-and-outlier-index}. If  one constraint has an $\mathcal{O}_j$ that is significantly higher compared to the others', that constraint is outlying. This is the case for CV SN, as shown in first row of Table \ref{tab:IOI-and-outlier-index}. The second row shows that removing CV SN from the constraint set reduces the IOI and most $\mathcal{O}_j$'s (except for H0LiCOW), indicating that most other constraints are, overall, consistent with each other. Our numerical analysis also indicates that the second most outlying constraint is H0LiCOW, this can be seen (1) in the second row of Table \ref{tab:IOI-and-outlier-index}, where H0LiCOW now has the highest $\mathcal{O}_j$ when CV SN is removed from the constraint list; and (2) in the third row, where the multi-dataset IOI further drops when H0LiCOW is removed in addition to CV SN and all $\mathcal{O}_j$ are relatively small. It is important for the constraints to be obtained from different types of observations, to avoid the possibility that some or most of them suffer from the same type of systematic effects. Planck and WMAP are both CMB observations, and may have similar systematic effects. But after removing WMAP from the constraint set, CV SN still has the highest $\mathcal{O}_j$; as shown in the fourth row. 

Both the constraint comparison shown in Figure~\ref{fig:all-H0-Om-constraints} and this numerical analysis point to our conclusion that the local measurement is the strongest driver of the Hubble tension, followed by the time-delay strong-lensing result. Given the fact the most other constraints are overall consistent with each other, this favors the possibility that some previously unseen systematic effects exist in the two most outlying constraints. However, more independent observations are needed in the future for a more decisive conclusion. 

Before we proceed, it is worth pointing out that Planck also has a relatively high $\mathcal{O}_j$ as seen from the first row in Table \ref{tab:IOI-and-outlier-index}, despite the fact that the Planck constraint overlaps with the white circle well (although not perfectly). But Planck's relatively high $\mathcal{O}_j$ is mainly driven by CV SN and H0LiCOW, which can been seen from the fact that Planck's $\mathcal{O}_j$ drops significantly after CV SN alone or both CV SN and H0LiCOW are removed from the comparison; see rows 2 and 3 in Table \ref{tab:IOI-and-outlier-index}. As we will further discuss later, there is some tension within the two most precise local measurements of $H_0$ \citep{Freedman-etal-2020}, and a recent analysis of time-delay strong-lensing gives a larger uncertainty and somehow lower value of $H_0$ \cite{Birrer-2020}. Using either alternative to these two constraints indeed significantly lowers most of the $\mathcal{O}_j$'s; see rows six and seven in Table \ref{tab:IOI-and-outlier-index}. On the other hand, since Planck is the strongest constraint, if it is particularly incompatible with most of other constraints, removing it from the analysis should lead to a even more significant drop of $\mathcal{O}_j$'s. But this is not the case. As we can see from the fifth row of Table \ref{tab:IOI-and-outlier-index}, a few $\mathcal{O}_j$'s still remain relatively high after Planck is removed from the analysis. This reflects the fact, mentioned earlier, that CV SN tends to overlap with different constraints in different parameter regions. 

We however would like to make clear that the consistency between Planck and most other observations found here does not imply a resolution or denial of the ``$\sigma_8$ tension.'' The Planck result has been shown to be in conflict with the inferred $\sigma_8$ (or some other combinations of $\sigma_8$ and $\Omega_{\rm{m}}$) from most LSS observations \citep{Wibking-etal-2019,Joudaki-etal-2019,2019-Boruah-etal} [though not all \citep{Hamana-etal-2019}]. Investigating the $\sigma_8$ tension is beyond the scope of this work. Nonetheless, we find that the Planck constraint does not on its own drive the current $H_0$ tension.

\section{Nonstandard cosmology} \label{sec:nonstandard}
Our analysis, above, examines the consistency of current constraints assuming a spatially flat $\Lambda$CDM model with no new physics. We emphasize that the above discussions are not rejecting the possibility of a nonstandard model that may reconcile all constraints considered. It is important to study nonstandard physics to see if the tension can be resolved. Our investigation is to serve as a complementary approach to the studies of nonstandard physics, and to provide a more comprehensive view of the current tension by including $\Omega_m$ rather than looking at the tension in terms of $H_0$ alone. A plethora of models have been proposed in the literature. While a detailed discussion of each model is beyond our scope here, we provide a general discussion of how high-$z$, mid-$z$, and low-$z$ nonstandard models as well as local environmental factors can alter constraints. It is important to note that any proposed model can change only a subset of constraints. This is another advantage of comparing constraints individually instead of comparing combinations of subsets of them: it is easier to see which constraints can be changed in a proposed model. In Figure~\ref{fig:all-H0-Om-constraints}, we label each constraint by ``H'' (high-$z$), ``M'' (mid-$z$), ``L'' (low-$z$) or ``E'' (local/ environmental) according to whether it can be changed by the corresponding models/proposals. (The H0LiCOW technique is relatively insensitive to the underlying cosmological model if the redshifts of the systems are not too high, so we leave it without a label.) Such a classification of the model dependences of each constraint is, to our knowledge, the first in the literature, which also helps navigate the search of nonstandard-physics resolution. We discuss below whether those models/proposals can reconcile all the constraints in tension.

{\bf High-$z$ models} (H) here refer to those have some non-standard physics before or around recombination, but reduce to a $\Lambda$CDM universe thereafter (e.g., by $z\sim100$). Examples are early-time dark energy Debates are ongoing about whether large-scale structure disfavors early dark energy models \citep{Hill-etal-2020,2020-Smith-etal}. \citep{Poulin-etal-2019-EDE}, self-interacting neutrinos \citep{Kreisch-etal-2019-self-interacting-neutrinos}  and primordial magnetic fields \citet{Pososian-Jedamzik2020}. Models of this category usually inject some extra energy before recombination, making the baryon-photon plasma sound horizon smaller. To compensate for this change and to match the observed angular size of the sound horizon, the Hubble constant needs to be higher than the CMB-inferred value in the $\Lambda$CDM model. Also, there is a ``theoretical correlation'' between the late-time BAO and the CMB observations in the sense that they both involve the calculation of the baryon-acoustic-oscillation sound horizon (although at two different epochs)\footnote{This does not mean the CMB and BAO are observationally dependent. We did not use any prior information from CMB in the BAO constraints, but rather used BBN to constrain $\Omega_bh^2$.}. It is therefore also possible for high-$z$ models to reconcile the late-time BAO constraints with CV SN. However, since these models reduce to a $\Lambda$CDM universe after recombination, they cannot change the constraints from late-time observations, especially those from $\gamma$-ray, CC, cosmic age and SN P. There would still be some tension remaining in the $H_0$-$\Omega_{\rm{m}}$ space for CV SN and those late-time constraints. For example, the 2-$\sigma$ contour of the $\gamma$-ray constraint does not overlap with the region of ($H_0\sim~74$\,km/s/Mpc, $\Omega_{\rm{m}}~0.3$). In addition, if any star with an age $\gtrsim13$\,Gyr is confirmed, it will disfavor High-$z$ models as solutions to the tension.

{\bf Mid-$z$ models} (M) refer to those have some non-standard physics after recombination but before $z\sim6$. An example is fractional decaying dark matter with a lifetime $\lesssim0.5$\,Gyr \citep{Vattis-Kyriakos-Abraham-2019-decayingDM}. Mid-$z$ models cannot change late-time constraints like $\gamma$-ray and CC. They might be able to change and loosen the cosmic-age bounds, but this requires the cosmic expansion to be slower during the mid-redshift range (compared to a standard evolution of $H$) to compensate for the rising of today's $H_0$. If this decrease of the earlier cosmic expansion extended to the very early universe (before recombination), it would conflict with the CMB and late-time BAO observations, as slower cosmic expansion before recombination makes the sound horizon even larger. 

{\bf Low-$z$ models} (L) refer to those have some non-standard physics during a redshift range relevant for most of the late-time observations considered here (i.e. $z\lesssim6$), but reduce to a standard cosmological model at higher redshifts. Recent examples are interacting dark energy \citep{Pan-etal-2019-interacting-DS}, a rolling scalar field \citep{Agrawal-etal-2019-Swampland-H0} and nonlocal modified gravity \citep{Belgacem-etal-2018-nonlocal-MG}. Most of the constraints here would be affected in some way by late-time evolution; however, it has been argued that late-time models are not able to reconcile the Hubble constant tension \citep{Aylor-etal-2019,Evslin-etal-2018-price-to-shift-H0}. Observations such as BAO and SNIa that probe the late-time background evolution can further constrain Low-$z$ models. 

While the above models can alter many of the constraints, some proposals suggest some {\bf local/environmental factors ($z\lesssim0.03$)} (E) can bias the local determinations. Local factors do not pose a problem to the standard $\Lambda$CDM model at large scales, but instead point to the need for a more detailed description of our local environment to account for such a systematic effect that can shift all local measurements in the same way. An example is a local underdense region \citep{2019-Shanks-etal-local-voids1,Lombriser-2019-H0-local-inhomogeneity}. Recent studies have shown observational evidence supporting a small-scale local underdense region \citep{Boehringer-etal-2019,2018-Pustilnik-etal-voids}. While it remains debated to what extend this may alleviate the $H_0$ tension, it has been strongly argued that it is unlikely to play a substantial role and may bias the local measurement at most by $1\%$ \citep{2019-Kenworthy-etal}. 

What puts another challenge to environmental-factor explanations is the H0LiCOW result. This is because the H0LiCOW technique is not a local measurement\footnote{By local measurements, we refer to methods that are \emph{only} based on the Hubble-Lema\^itre law and small-$z$ ($z\lesssim0.1$) observations.}, although it is relatively insensitive to the underlying cosmological model. It involves observations at higher redshifts than those used for the local measurements and a local structure, like a void, can barely shift the H0LiCOW result. Thus separate considerations are needed to explain the high H0LiCOW result. Incidentally, some authors have suggested that the H0LiCOW technique can be affected by the mass-sheet degeneracy and the issue may be more complicated than is assumed \citep{2019Gomer-etal,Blum-etal-2020}. Indeed, by relaxing the strong assumption on the lens density profile and adding more external data, \citet{Birrer-2020} found that the uncertainty of $H_0$ is noticeably larger, with a lower mean value. As a consequence, using this TDSCOSMO result indeed of H0LiCOW in the numerical analysis makes CV SN more outlying and strengthens our conclusion; see the seventh row of Table \ref{tab:IOI-and-outlier-index}.

Comparing different local measurements may provide hints for environmental effects if any, but even some local measurements from different methods somewhat disagree. The recent TRGB-based local measurement reported $H_0=69.8\pm1.9$\,km/s/Mpc \citep{Freedman-etal-2019}, which is more consistent with Planck than with CV SN. Other authors revisited this result and found a higher value, $H_0=72.4\pm2.0$\,km/s/Mpc \citep{Yuan-etal-2019-TRGB-local}. But \citet{Freedman-etal-2020} later showed more details of their analysis and confirmed their relatively low local measurement with $H_0=69.6\pm1.9$\,km/s/Mpc. Using other methods to calibrate SNIae overall have relatively large uncertainties. Calibration using surface brightness fluctuations (SBF) gives $H_0=70.50\pm2.37(\rm{stat})\pm3.38(\rm{sys})$\,km/s/Mpc \citep{Khetan-etal-2020}, more consistent with the TRGB-based measurement. The Mira-based method reported $H_0=73.3\pm3.9$\,km/s/Mpc \citep{2019-Huang-etal-Mira-H0}, more consistent with the CV SN result. Those above local measurements are based on Type Ia supernovae (using different calibrations); it is possible that they could share some common systematic effects. For example, the local $H_0$ may be overestimated if the Hubble-flow SNIae are intrinsically brighter than those in the calibration samples due to a difference in environmental age \citep{Rigault-etal-2018,Kang-etal-2020} [but also see \citet{Rose-etal-2019} for discussion]. Or, there might be other nonstandard physical reasons such as screened fifth forces that affect the calibration of SNIae  \citep{Desmond-Jain-Sakstei-2019-fifth-force}.

Replacing SNIae with other secondary distance indicators can test the possible unaccounted-for systematic errors on the SNIa side. For example, it was recently reported that $H_0=76.0\pm2.5$\,km/s/Mpc using the Tully-Fisher relation (TFR) \citep{2020-Kourkchi-etal}. A similar result was obtained using the baryonic Tully-Fisher relation \citep{2020-Schombert-etal}. With a larger uncertainty, \citet{2020-Jaeger-etal} reported $H_0=75.8^{+5.2}_{-4.9}$\,km/s/Mpc using Type II supernovae (SN II). These works however used Cepheid variables or TRGBs in calibration and thus are not totally independent local measurements. As such, differences in the TRGB calibrations lead to complicated agreements or discrepancies among different local measurements \citep{2020-Kourkchi-etal}. There are local determinations that do \emph{not} require the distance ladder or SNIae, such as the MCP result \citep{Pesce-etal-2020} and the standard siren multimessenger method \citep{2019-GW-BNS-update}. The latest MCP result is $73.9\pm3.0$\,km/s/Mpc. This a relatively weak constraint and is consistent with most of the other constraints considered here. The gravitational wave multimessenger constraint is still too weak to play a significant role.

\section{Discussion} \label{sec:discussion}
We present these results as an exploration of the range of possibilities for approaching tensions among these disparate datasets. It should be noted that our numerical analysis does not include the cosmic-age bounds, since single-sided constraints cannot be considered in the moment-based IOI formalism. However, considering these bounds can only strengthen our conclusion, since they overall put pressure on the regions where CV SN overlaps with SNIa and DES, as well as the region where it overlaps with BAO Gal. As mentioned earlier, the recent TRGB-based determination of $H_0$ is lower than that of CV SN \citep{Freedman-etal-2020}. If we replace CV SN with this result and redo our numerical analysis, IOI and most $\mathcal{O}_j$'s (except for H0LiCOW) are significantly reduced; see the sixth row in Table \ref{tab:IOI-and-outlier-index}. We can also see the consistency between the TRGB result and most other constraints in the left panel of Figure~\ref{fig:other-localH0} in Appendix \ref{appendix:otherlocal}. In addition to replacing CV SN with TRGB, removing H0LiCOW further enhances the consistency among constraints; see the seventh row in Table \ref{tab:IOI-and-outlier-index}. Recently, \citet{2019-Ivanov-etal,Philcox-etal-2020} improved the analysis of the galaxy power spectrum (embedding BAO) at $z_{\rm{eff}}=0.38$ and $0.61$ using different methods that accounts for non-linear clustering and a range of other effects. Their constraints on the $H_0$-$\Omega_{\rm{m}}$ plane also fall into the guiding circle in Figure~\ref{fig:all-H0-Om-constraints}.

A caveat in our investigation is that all observations are treated upon the same footing in the sense that not any one of them is considered as more reliable than any other. In practice, some results may be more or less vulnerable to systematic effects. In particular, while promising, the CC constraint is a relatively new technique and studies of additional systematic errors are ongoing \citep{2020-Moresco-Raul-Verde}. We however note that our CC constraint has included the estimated additional systematic errors considered in \citet{2020-Moresco-Raul-Verde}; see our Appendix \ref{appendix:additional-sys-cc}. As mentioned earlier, a better understanding and constraining of the lens model is required before a robust constraint on $H_0$ from strong gravitational lensing can be achieved \citep{Birrer-2020}. In addition, knowing the cosmic age is promising to constitute a strong test to high-$z$ (and likely also mid-$z$) models. In fact, any objects or methods that confirm a cosmic age $\gtrsim13.5$\,Gyr since $z=100$ (e.g., if the aforementioned estimated age of J18082002-5104378 A, $13.535\pm0.002$\,Gyr, is confirmed) can rule out all high-$z$ solutions to the current $H_0$ tension.

\section{Conclusions} \label{sec:conclusion}
We have shown that it is very beneficial to compare constraints as individually as possible in the $H_0$-$\Omega_{\rm{m}}$ space when investigating the $H_0$ tension. It allows us to more robustly see how different constraints behave in the standard $\Lambda$CDM model as well as to more easily tell whether a nonstandard model can reconcile all constraints.

We have performed a systematic comparison of various constraints from different observations in the $H_0$-$\Omega_{\rm{m}}$ space based on the standard $\Lambda$CDM model. Most constraints consistently overlap along different degeneracy directions on some common region in the $H_0$-$\Omega_{\rm{m}}$ plane centered around the general vicinity of (68.5,0.3). The fact that the Cepheid-based local determination of $H_0$ does not overlap with such a common region suggests that the main driver of the tension may be supposed to be the local measurement(s). The time-delay strong-lensing result only marginally overlaps with that region, making it the next most outlying constraint. While we do not reject the possibility that some nonstandard physics may resolve the current Hubble constant tension, we found that it is difficult for high-$z$, mid-$z$, or low-$z$ nonstandard evolution models or local environmental factors to reconcile the constraints of all the considered observations as they stand. Confirming the results from $\gamma$-ray attenuation and cosmic chronometers as well as the lower limit of cosmic age can rule out high- and mid-$z$ models that try to resolve the current $H_0$ tension. Standard rulers and candles will continue to put pressure on low-$z$ models. Solutions that alter the local determination need a separate explanation for the high time-delay strong-lensing result. 

In the future, there will be more and more independent methods to constrain the cosmic evolution. For instance, observing the redshift drift will allow us to directly detect the real-time cosmic expansion \citep{Loeb-1998}. BAO constraints will be improved from the line-intensity mapping of emission from star-forming galaxies \citep{Bernal-etal-2019-BAO-intensity-map} and the next-generation galaxy surveys \citep{2019-Bengaly-etal-BAO-galaxysurvey}. The drop-off in the abundance of $\gtrsim45M_{\odot}$ black holes can be used to probe cosmic expansion by making binary black hole mergers ``standardizable'' \citep{2019-Farr-Fishbach-Ye-Holz}. Velocity-induced acoustic oscillations, a standard ruler that can be seen in 21-cm power spectrum, provide a way to probe the background evolution at cosmic dawn \citep{2019-Munoz-VAO}. And standard siren multimessengers provide another way to measure $H_0$. Comparing them all in the $H_0$-$\Omega_{\rm{m}}$ space can help us to more easily discover the fundamental cause of the Hubble constant tension.

It is exciting to see whether the current Hubble constant tension is leading us to another new understanding of the universe. We hope that this analysis motivates a new way of considering the various cosmological constraints and a different perspective on viewing such a tension.

\acknowledgments
We thank Alberto Dom\'inguez D\'iaz and Radoslaw Jan Wojtak for providing the MCMC chain of the analysis to the $\gamma$-ray attenuation data,  Simon Birrer, Mustapha Ishak, Khaled Said, Brian Schmidt, Ji Yao and Matias Zaldarriaga for useful comments and suggestions, Adam Mantz and Sebastian Bocquet for answering questions on some particular cosmological constraints, and Adam Riess for helpful discussions and references. We especially thank Michele Moresco for the advice on how to consider additional systematic errors in the current cosmic chronometer data. KJM acknowledges the Simons Emmy Noether Fellowship, which has supported her residence at the Perimeter Institute while the later stages of this work were carried out. We also thank an anonymous referee for feedback and suggestions that have improved this work.

\appendix

\section{Likelihood for lens system DES J0408}\label{appendix:lens7-likelihood}
The latest joint H0LiCOW analysis only includes six lens systems \citep{Wong-etal-2019-joint-stronglensing}, while there has been an individual analysis of a seventh lens system, DES J0408 \citep{Shajib-etal-2020}. The analysis of DES J0408 alone gives $H_0$\,=\,$74.0^{+2.7}_{-3.0}$\,km/s/Mpc in the standard $\Lambda$CDM model. While the chains of the time-delay distance $D_{\Delta t}$ and the lens angular diameter distance $D_{\rm{L}}$ for this system are not publicly available, we approximated the likelihoods of $D_{\Delta t}$ and $D_{\rm{L}}$ for lens DES J0408 by a skewed Gaussian distribution,
\begin{equation}\label{eq:skewed-Gaussian}
    \mathcal{L}(x)\propto\left[1+\rm{erf}(s\frac{x-x_0}{\sigma})\right]\exp\left(-\frac{(x-x_0)^2}{2\sigma^2(1+|s|)}\right)\,,
\end{equation}
where $x$ stands for $D_{\Delta t}$ or $D_{\rm{L}}$ in Mpc. The parameters $(x_0,\,\sigma,\,s)$ are different for $D_{\Delta t}$ and $D_{\rm{L}}$. They are chosen to reproduce the median statistics of $D_{\Delta t}$\,=\,$3382^{+146}_{-115}$\,Mpc and $D_{\rm{L}}$\,=\,$1711^{+376}_{-280}$\,Mpc reported in \citep{Shajib-etal-2020} and are $(3231,\,133.5,\,1.17)$ for $D_{\Delta t}$ and $(1368,\,331.4,\,1.37)$ for $D_{\rm{L}}$. We have ignored the correlation between $D_{\Delta t}$ and $D_{\rm{L}}$. Using the same priors on $H_0$ and $\Omega_{\rm{m}}$ in \citet{Shajib-etal-2020}, we closely reproduced the posterior $H_0$\,=\,$74.0^{+2.7}_{-3.1}$\,km/s/Mpc for lens DES J0408-5354 after marginalizing over $\Omega_{\rm{m}}$. The two likelihoods of $D_{\Delta t}$ or $D_{\rm{L}}$ are then added to the joint H0LiCOW likelihood. The joint marginalized constraint on $H_0$ is $73.7^{+1.5}_{-1.6}$\,km/s/Mpc for all the seven lens systems.

\section{Estimating systematic errors for cosmic chronometer data}\label{appendix:additional-sys-cc}
Systematic effects for the cosmic chronometer measurements have been extensively considered in \citet{2020-Moresco-Raul-Verde}. While their previous works have considered systematic errors due to the young component contamination and the star formation history dependence \citep{2017-Moresco-Marulli-CC,2016-Moresco-etal}, \citet{2020-Moresco-Raul-Verde} focuses on additional systematic errors due to the initial mass function (IMF) and stellar population synthesis (SPS) model. We added these to the covariance matrice of the current CC data to better represent the full range of errors. Specifically, we utilized columns 2 and 5 in \citet{2020-Moresco-Raul-Verde}'s table 3 for these two systematic errors. When using column 5, we have assumed the most discordant SPS model(s) can be discarded. This is fairly reasonable because for the SPS models available in the literature, \citet{2020-Moresco-Raul-Verde} found that one of them is significantly different from the others while others are in fair agreement. Therefore, when compared against high signal-to-noise and high spectral data to see which SPS model(s) is (are) better, which can be achieved by using the new high-resolution instruments, e.g., X-Shooter or VIMOS/HR-Red. We interpolated the data listed in their table 3 to get the error budget of the current CC data at each redshift due to these two extra sources. We then built the covariance matrices, $\rm{Cov}^{\rm{IMF}}_{i,j}$ and $\rm{Cov}_{i,j}^{\rm{SPS}}$, according to their Eq.\,9, i.e.,
\begin{equation}\label{eq:Cov-CC-stimated}
    {\rm{Cov}}^{\rm{X}}_{i,j}=\widehat{\eta^{\rm{X}}}(z_i)H(z_i)\widehat{\eta^{\rm{X}}}(z_j)H(z_j)\,,
\end{equation}
where $\widehat{\eta^{\rm{X}}}(z)$'s are obtained by interpolation with the data provided in their table 3, $H(z_i)$'s are CC measurements at different redshifts. At last, we added $\rm{Cov}^{\rm{IMF}}_{i,j}$ and $\rm{Cov}_{i,j}^{\rm{SPS}}$ to the covariance matrix of the current CC data and performed a likelihood analysis with the standard $\Lambda$CDM model to obtain the CC constraint in the $H_0$-$\Omega_{\rm{m}}$ plane. We note that the estimated $\rm{Cov}^{\rm{IMF}}_{i,j}$ and $\rm{Cov}_{i,j}^{\rm{SPS}}$ according to Eq.\,(\ref{eq:Cov-CC-stimated}) are correlated at all redshifts. But as pointed out in \citet{2020-Moresco-Raul-Verde} no model systematically overpredicts or underpredicts the Hubble parameter with respect to another, so these systematic errors are not quite correlated at different redshift bins. We however choose to include the correlation of those extra systematic errors and therefore the constraint given here should be viewed as a conservative estimate. Detailed consideration of additional systematic errors for CC constraints are works in progress \citep{2020-Moresco-Raul-Verde}.

\section{Remarks on the combination of BAO and BBN}\label{appendix:remarks-BAO-BBN}
In the context of the standard $\Lambda$CDM model\footnote{By the standard $\Lambda$CDM model, we refer to one with the number of effective relativistic particles, neutrino mass, spatial curvature, etc., fixed to the standard values \citep{2018-Planck-cosmo-params}. }, combining BAO with the baryon density parameter inferred from BBN has become a standard procedure to provide a statistically independent constraint on $H_0$ \citep{Cuceu-etal-2019,2019-Schoneberg-etal}. However, when the baryon acoustic sound horizon scale at the drag epoch $r_{\rm{d}}$ is treated as a free parameter, BAO alone is not able to constrain $H_0$ \citep{Aylor-etal-2019}. This is because $r_{\rm{d}}$ is degenerate with $H_0$ and $r_{\rm{d}}$ needs to be known to constrain $H_0$ from BAO. This $r_{\rm{d}}$ can be modeled once a cosmological model is assumed such as the standard $\Lambda$CDM model in our main analysis. The calculation of $r_{\rm{d}}$ only involves physics before the drag epoch [$z_{\rm{d}}\sim1060$ \citep{2018-Planck-cosmo-params}]; as such, $r_{\rm{d}}$ is often said to be ``calibrated at early time.'' One may argue that $r_d$ can also be calibrated (or, more properly, inferred) if we treat it as a free parameter and combine BAO with some local measurement of $H_0$ \citep{Addison-etal-2018}, since knowing $H_0$ gives us $r_{\rm{d}}$ from BAO. However, inferring $r_{\rm{d}}$ in this way already assumes the local measurement used to infer the sound horizon scale to be correct and that there is some beyond-the-standard-$\Lambda$CDM physics. his approach does not match our goal. This is because the first part of this work is to investigate which method(s) gives the most incompatible constraint(s) in the $H_0$-$\Omega_{\rm{m}}$ plane to the others when the standard $\Lambda$CDM model is assumed. It would be an unfair comparison to relax the the assumption of the standard $\Lambda$CDM model for BAO while keeping this assumption for other constraints. However, we provided a general discussion on nonstandard models in Sec.\,\ref{sec:nonstandard}.

BBN provides a prior on the reduced baryon fraction $\Omega_{\rm{b}}h^2$, about which there is no substantial disagreement from different observations based on completely different physics, e.g., CMB anisotropies \citep{2018-Planck-cosmo-params}. The $\Omega_{\rm{b}}h^2$ values inferred from various different effects on CMB anisotropies are also consistent with each other \citep{2020a-Motloch}. This gives us some confidence of the robustness of this prior on $\Omega_{\rm{b}}h^2$. It has also been pointed out in \citet{Verde-etal-2019} that, for a nonstandard model proposed to resolve the $H_0$ tension, ``any change in background parameters (physical densities) should be mostly via $H_0$ and not via the density parameters themselves.'' Furthermore, \citet{Cuceu-etal-2019,2019-Schoneberg-etal} showed that the current BAO results only have a weak dependence on $\Omega_{\rm{b}}h^2$, and we have adopted an even weaker BBN prior on $\Omega_{\rm{b}}h^2=0.0222\pm0.0005$ than those adopted in \citet{Cuceu-etal-2019,2019-Schoneberg-etal}. 

One may worry that the same BBN constraint is included in both BAOs, which may render the two BAO results not independent. To dispel this latter worry, we perform a numerical analysis with both BAOs combined so that BBN is used only once. Doing so does not qualitatively change our conclusions; see the last row in Table \ref{tab:IOI-and-outlier-index}.

\section{Other local measurements and large-scale-structure constraints}\label{appendix:otherlocal}
\begin{figure}[h]
    \centering
    \includegraphics[width=0.495\textwidth]{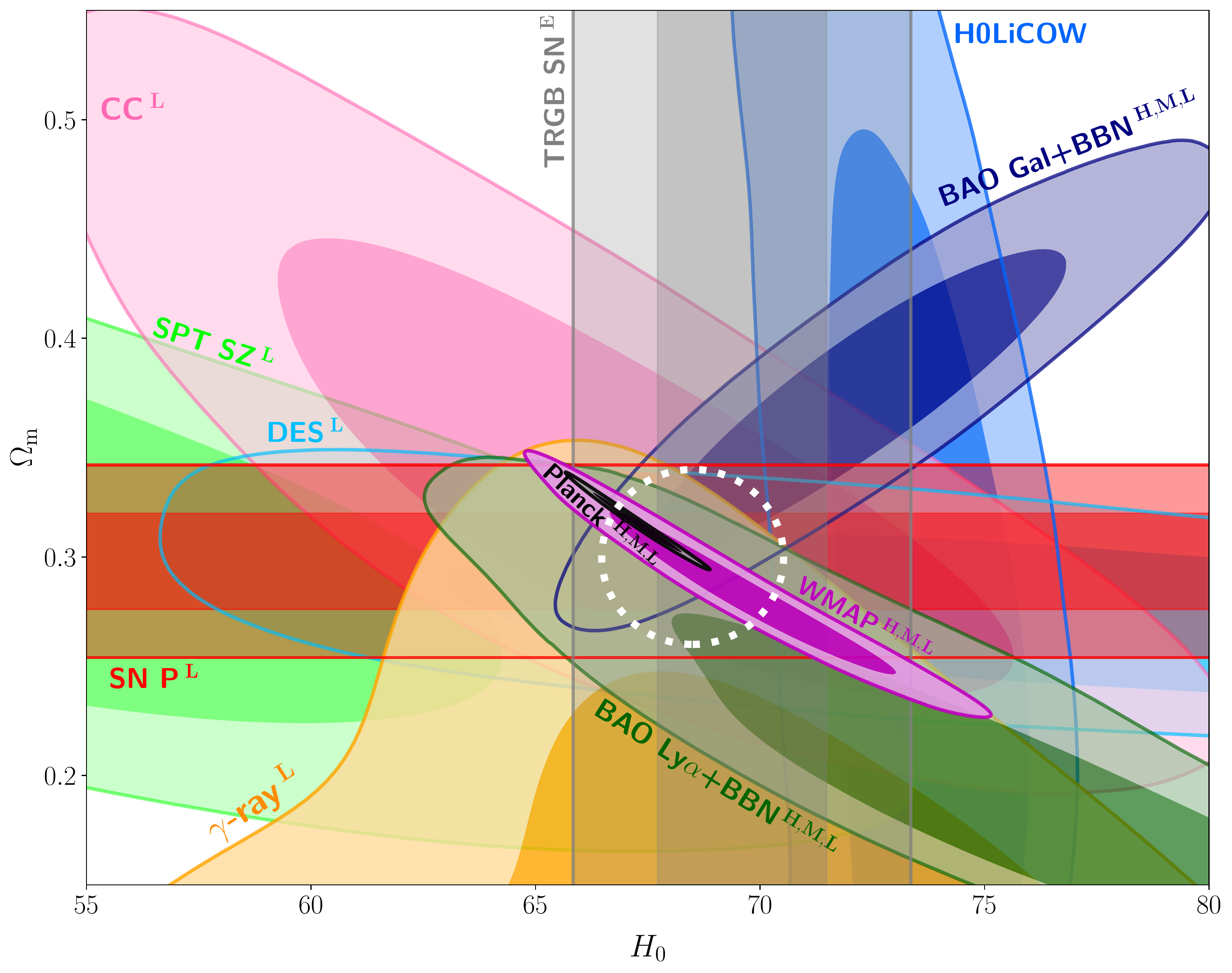}
    \includegraphics[width=0.495\textwidth]{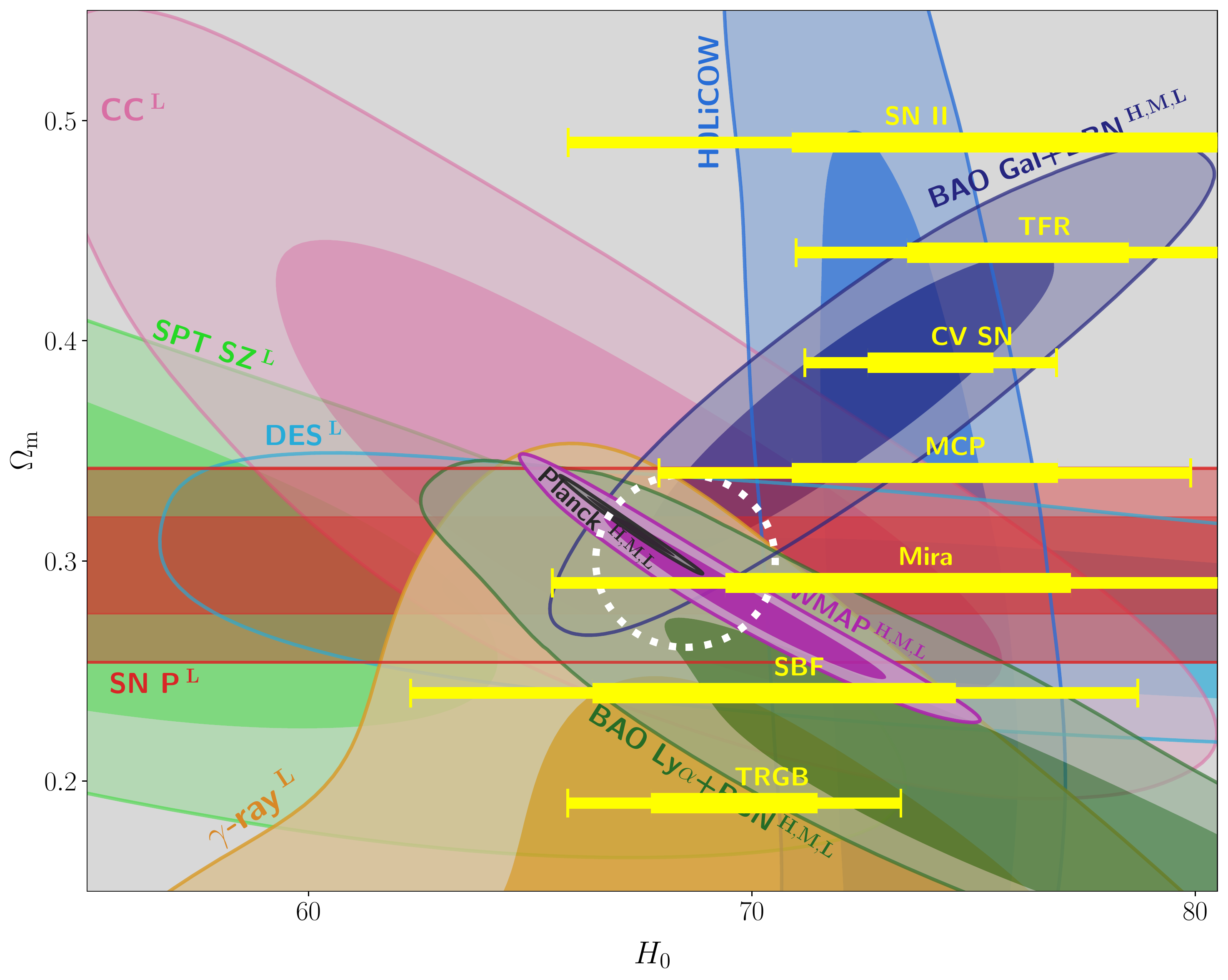}
    \caption{Different constraints in the $H_0$ and $\Omega_{\rm{m}}$ space based on a flat $\Lambda$CDM model, as in Figure~\ref{fig:all-H0-Om-constraints}. As above, dark and light contours show the $68\%$ and $95\%$ confidence regions of each posterior. Here, the cosmic age bound is omitted. In the left panel, we plot the TRGB constraint instead of CV SN, and on the right we plot all the quoted local measurements of $H_0$ in Sec.\,\ref{sec:nonstandard}. Those error bars in the right panel should be interpreted as vertical bands with the thick and thin lines representing 1-$\sigma$ and 2-$\sigma$ confidence ranges, respectively. Local measurements tend to give a higher $H_0$ value than that indicated by the white circle. However, we note that they are not all independent from each other, and their favored ranges depend on the calibration adopted; see the end of Sec.\,\ref{sec:nonstandard} for a discussion. } In both panels, we included the constraints from SPT SZ clustering (lime), for which we restrict that $0.0006\leq\Omega_\nu h^2\leq0.0009$ for the standard $\Lambda$CDM since the chains obtained from \href{https://pole.uchicago.edu/public/data/sptsz-clusters/}{https://pole.uchicago.edu}] are based on a $\nu\Lambda$CDM model with a large range of $\Omega_\nu h^2$. 
    \label{fig:other-localH0}
\end{figure}

Here, we include versions of Figure~\ref{fig:all-H0-Om-constraints}, plotting other local measurements (TRGB and MCP) in place of the CV SN constraint. We also plot SPT SZ in both panels. In our numerical analysis in the previous sections, CV SN is replaced by TRGB in some cases to see how this alternative distance ladder measurement compares to other constraints, but MCP, as an independent local measurement, is always included. In the left panel of Figure~\ref{fig:other-localH0} we plot the TRGB constraint, and in the right panel we plot all the quoted local measurements. The thick and thin bars in the right panel represent 1-$\sigma$ and 2-$\sigma$ confidence ranges, respectively, and should be interpreted as vertical bands. We note that not all local measurements are independent so in our numerical analyses we did not include all of them at the same time but alternatively used the CV SN and TRGB results; see Sec.\,\ref{sec:nonstandard} for a discussion.   As can be seen in right panel, the independent MCP constraint prefers a high $H_0$ value but is broader than the CV SN allowed region, and therefore has some overlap with the guiding white circle (with the $2$-$\sigma$ contour). The TRGB constraint is in better agreement with other constraints, overlapping the white guiding circle substantially. Quantitatively, the level of concordance can be seen in Tables \ref{tab:IOI-and-outlier-index} discussions thereof. It is worth mentioning that the joint analysis of 6dFGS and SDSS peculiar velocity results indicates some indirect preference for $H_0>70$\,km/s/Mpc and (when combined with other probes as needed) may also provide an interesting constraint on the $H_0$-$\Omega_{\rm{m}}$ plane \citep{2020-said-etal}. 

\bibliography{LCDM2N}
\bibliographystyle{aasjournal}

\end{document}